\documentclass{revtex4}
\usepackage{mathbbold}
\usepackage{amsfonts}
\usepackage{amsmath}
\usepackage{amssymb}
\usepackage{charter}
\usepackage{graphicx}

\begin{document}

\title{On the optical fields propagation in realistic environments}
\author{Xue-xiang Xu$^{\dag }$}
\affiliation{Center for Quantum Science and Technology, Jiangxi Normal University,
Nanchang 330022, China\\
$^{\dag }$Corresponding author: xuxuexiang@jxnu.edu.cn }

\begin{abstract}
Evolution formulas\ of the density operator, the photon number distribution,
and the Wigner function are derived for the problem on the optical fields
propagation in realistic environments. The method of deriving these formulas
is novel and the results are very useful for quantum optics and quantum
statistics.
\end{abstract}

\maketitle

\section{Introduction}

In quantum optics and quantum statistical mechanics, people often come
across such problems: (1) when a system is immersed in a realistic
environment; or (2) a signal (a quantum state) passes through a quantum
channel. Among them, the decay of the radiation field inside a cavity plays
an important role in many realistic problems. In general, damping of the
radiation field is described by its interaction with a reservoir with a
large number of degrees of freedom. However, we are interested in the
evolution of the variables associated with the optical field only. This
requires us to obtain all relevant properties for the optical field of
interest only after tracing over the reservoir variables \cite{1,2}.

Actually, the decay of the radiation field inside a cavity is an problem on
the dynamics of a open quantum system \cite{3,4}. One can use the Liouville
equation (a master equation) to describe the dynamical evolution of the
density matrix of the optical field. That is, the decay (or decoherence) due
to the interaction between a system and its environment can be described by
a Liouville (super) operator. However, the Liouville (super) operator
describes a nonunitary time evolution, which cannot trivially be integrated
through standard Lie-algebra techniques. Analytical solutions of the
Liouville equation can be obtain upon resorting to quasiprobability
representations of the density matrix \cite{5}, or upon evaluating
eigenvalues and eigenvectors (i.e. eigen density matrices) of the Liouville
(super) operator \cite{6}, or even upon using\ group-theoretical approach
\cite{7}, or by virtue of the thermo-entangled state representation \cite{8}.

Followed by above works, we also pay our attention to study the optical
fields propagation in realistic environment in this paper. The physical
problem is abstracted into a mathematical model. Then using our technique on
the quantum operators and the quantum states, we cleverly deduce some
formulas for the propagated optical fields. These formulas are very useful
to quantum optics and quantum statistics.

The manuscript is organized as follows. We start in Sec. II by introducing
the theoretical model and abstracting two formalism. That is, we imagine the
fact of the optical fields propagation in realistic environment as a
fictitious beam splitter (BS) model and construct the correspondences
between the optical channel formalism and the BS formalism. In Sec. III, as
the section of the BS formalism, we derive the density operator of the
output optical fields by using the Weyl expansion of the density operator in
the characteristic function (CF) formalism. Moreover, the formula of the
photon number distribution (PND) and the Wigner function (WF) for the output
state are derived in detail. In Sec. IV, as the section of the
optical-channel formalism, we obtian the time evolution formula of the
density operator, the PND and the WF. As the application of these formulas,
we use a single-photon-addition coherent state as the initial state and
discuss its evolution in thermal channel in Sec.V. Our conclusions are
summarized in the last section.

\section{Physical and mathematical model}

The fact that a optical field propagation in the realistic environment (in
optical-channel formalism) can be simulated as the interaction of a
fictitious BS (in the BS formalism). The initial optical field $\rho \left(
0\right) $ (denoted by the input quantum state $\rho _{in}$) in the main
mode $a$ and the environment (also reservoir or channel, denoted by a
general environment state $\rho _{b}$) in the auxiliary mode $b$ are
interacted by a BS (described by the operator $B$). The output quantum state
$\rho _{out}$ (i.e. the final optical field $\rho \left( \tau \right) $) can
be given by making partial trace over the ancillary mode $b$,
\begin{equation}
\rho _{out}=\mathrm{Tr}_{b}\left[ B\left( \rho _{in}\otimes \rho _{b}\right)
B^{\dag }\right] .  \label{1}
\end{equation}%
\begin{figure}[tbp]
\label{Fig1} \centering\includegraphics[width=0.8\columnwidth]{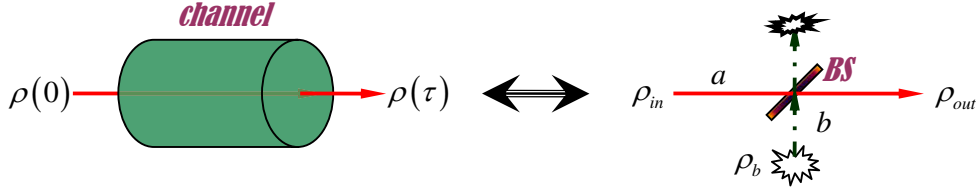}
\caption{(Colour online) Theoretical equivalent model. Optical fields
propagation in realistic environment, can be equivalent to the optical field
propagation in a optical channel formalism and also can be abstracted to a
fictitious beam-splitter formalism .The correspondences include $\protect%
\rho \left( 0\right) \Longleftrightarrow \protect\rho _{in}$; $\protect\rho %
\left( \protect\tau \right) \Longleftrightarrow \protect\rho _{out}$.}
\end{figure}

Our equivalent model ia shown in Fig.1. There exist the one-to-one
correspondence between the optical-channel formalism and the BS formalism.
In their respective formalism, we have
\begin{equation}
\left(
\begin{array}{c}
a\left( \tau \right) \\
b\left( \tau \right)%
\end{array}%
\right) =\left(
\begin{array}{cc}
e^{-\kappa \tau /2} & \sqrt{1-e^{-\kappa \tau }} \\
-\sqrt{1-e^{-\kappa \tau }} & e^{-\kappa \tau /2}%
\end{array}%
\right) \left(
\begin{array}{c}
a\left( 0\right) \\
b\left( 0\right)%
\end{array}%
\right) \Longleftrightarrow B\left(
\begin{array}{c}
a \\
b%
\end{array}%
\right) B^{\dag }=\left(
\begin{array}{cc}
t & r \\
-r & t%
\end{array}%
\right) \left(
\begin{array}{c}
a \\
b%
\end{array}%
\right) .  \label{2}
\end{equation}%
In optical-channel formalism, $a\left( \tau \right) $ is the attenuated
light mode, $b\left( \tau \right) $ is a flunctutation mode, and $\kappa $\
is a damping constant. While in the BS formalism, we defined the BS operator
$B=\exp \left( \theta \left( a^{\dag }b-ab^{\dag }\right) \right) $ in terms
of the creation (annihilation) operator $a^{\dag }$ ($a$) and $b^{\dag }$ ($%
b $) with the transmission coefficient $t=\cos \theta $ and $r=\sin \theta $%
. The relationship can be linked by $t=e^{-\kappa \tau /2}$, $r=\sqrt{\left(
1-e^{-\kappa \tau }\right) }$, and $\rho \left( 0\right) \Longleftrightarrow
\rho _{in}$; $\rho \left( \tau \right) \Longleftrightarrow \rho _{out}$. It
should be emphased that $t$ is the transmission coefficient of the BS and $%
\tau $ is the time of the optical field propagation in the channel.

As pointed in Ref.\cite{9}, the simple intuitive model is exact for
dissipation in Gaussian reservoirs. Hence, we take a squeezed thermal state
(a general Gaussian quantum state), i.e.%
\begin{equation}
\rho _{b}=S\left( \lambda \right) \rho _{th}\left( \bar{n}\right) S^{\dag
}\left( \lambda \right) ,  \label{3}
\end{equation}%
as the environment. This environment can be reduced the reservoirs of vacuum
($\bar{n}=\lambda =0$), thermal ($\lambda =0$), or squeezed vacuum ($\bar{n}%
=0$). By the way, $S\left( \lambda \right) =\exp \left[ \frac{\lambda }{2}%
\left( a^{\dagger 2}-a^{2}\right) \right] $ is the single-mode squeezed
operator with real squeezing parameter $r$. Moreover, $\rho _{_{th}}\left(
\bar{n}\right) =\sum_{n=0}^{\infty }\frac{\bar{n}^{n}}{\left( \bar{n}%
+1\right) ^{n+1}}\left\vert n\right\rangle \left\langle n\right\vert $\ is a
thermal state with the average photon number $\bar{n}$.

\section{Beam splitter formalism}

In this section, we shall derive the output density operator, the output
photon number distribution, and the output WF in the BS formalism.

\subsection{Density operator}

By using the Weyl expansion of the density operator \cite{10,11}, we can
express $\rho _{in}$ and $\rho _{b}$\ in the CF formalism%
\begin{equation}
\rho _{in}=\int \frac{d^{2}\alpha }{\pi }\chi _{in}\left( \alpha \right)
D_{a}\left( -\alpha \right) ,  \label{4}
\end{equation}%
and%
\begin{equation}
\rho _{b}=\int \frac{d^{2}\beta }{\pi }\chi _{b}\left( \beta \right)
D_{b}\left( -\beta \right)  \label{5}
\end{equation}%
where $D_{a}\left( \alpha \right) =\exp \left( \alpha a^{\dag }-\alpha
^{\ast }a\right) $ is the displacement operator in mode $a$, and $\chi
_{in}\left( \alpha \right) =$Tr$\left( \rho _{in}D_{a}\left( \alpha \right)
\right) $ is the CF of $\rho _{in}$. Similarly, $D_{b}\left( \beta \right)
=\exp \left( \beta b^{\dag }-\beta ^{\ast }b\right) $ is the displacement
operator in mode $b$, and $\chi _{b}\left( \beta \right) =$Tr$\left( \rho
_{b}D_{b}\left( \beta \right) \right) =\exp \left( -M\left\vert \beta
\right\vert ^{2}+N\beta ^{2}+N\beta ^{\ast }{}^{2}\right) $ is the CF of $%
\rho _{b}$ with $M=\left( \bar{n}+\frac{1}{2}\right) \allowbreak \cosh
2\lambda $ and $N=\allowbreak \frac{1}{2}\left( \bar{n}+\frac{1}{2}\right)
\sinh 2\lambda $.

Substituting Eqs. (\ref{4})\ and (\ref{5})\ into Eq.(\ref{1}) and using the
BS transformation relations, we have%
\begin{eqnarray}
\rho _{out} &=&\int \frac{d^{2}\alpha }{\pi }\int \frac{d^{2}\beta }{\pi }%
\chi _{in}\left( \alpha \right) \chi _{b}\left( \beta \right) D_{a}\left(
\left( r\beta -t\alpha \right) \right) \mathrm{Tr}_{b}\left[ D_{b}\left(
-\left( r\alpha +t\allowbreak \beta \right) \right) \right]  \notag \\
&=&\frac{1}{t^{2}}\int \frac{d^{2}\alpha }{\pi }\chi _{in}\left( \alpha
\right) D_{a}\left( -\frac{1}{t}\alpha \right) \exp \left( -X\left\vert
\alpha \right\vert ^{2}+Y\alpha ^{2}+Y\alpha ^{\ast 2}\right) .  \label{6}
\end{eqnarray}%
with $X=Mr^{2}/t^{2}$ and $Y=Nr^{2}/t^{2}$. In the above step, we have used
the relation $\mathrm{Tr}_{b}\left[ D_{b}\left( -\left( r\alpha
+t\allowbreak \beta \right) \right) \right] =\pi \delta ^{(2)}\left( -\left(
r\alpha +t\allowbreak \beta \right) \right) $ and $\delta ^{(2)}\left(
-\left( r\alpha +t\allowbreak \beta \right) \right) =\frac{1}{t^{2}}\delta
^{(2)}\left( -\left( \frac{r}{t}\alpha +\beta \right) \right) $ \cite{12}.
Eq.(\ref{6}) can also be written as%
\begin{equation}
\rho _{out}=\int \frac{d^{2}\alpha }{\pi }\chi _{in}\left( t\alpha \right)
D_{a}\left( -\alpha \right) \exp \left( -Mr^{2}\left\vert \alpha \right\vert
^{2}+Nr^{2}\alpha ^{2}+Nr^{2}\alpha ^{\ast 2}\right) .  \label{7}
\end{equation}%
Therefore, once the input CF $\chi _{in}\left( \alpha \right) $ is known,
then the density operator of the output optical field can be obtained by
performing the integration in Eqs. (\ref{6}) or (\ref{7}).

\subsection{Photon number distribution}

The PND is a key characteristic of every optical field. All interesting
states of the field are constructed as a combination of Fock states, and
different combinations have different quantum properties. Optical field
propagation at different time has different character by analyzing the PND.
Noticing the definition of the PND, i.e. $P_{out}\left( n\right)
=\left\langle n\right\vert \rho _{out}\left\vert n\right\rangle $ and using
Eq.(\ref{7}), we have%
\begin{eqnarray}
P_{out}\left( n\right) &=&\frac{1}{n!}\frac{d^{2n}}{dh^{n}ds^{n}}\int \frac{%
d^{2}\alpha }{\pi }\chi _{in}\left( \alpha t\right) \exp \left( -\left(
Mr^{2}+\frac{1}{2}\right) \left\vert \alpha \right\vert ^{2}\right)  \notag
\\
&&\times \exp \left( +Nr^{2}\alpha ^{2}+Nr^{2}\alpha ^{\ast 2}+\allowbreak
h\alpha ^{\ast }-s\alpha +hs\right) |_{s=h=0}  \label{8}
\end{eqnarray}%
where we have used that $D_{a}\left( -\alpha \right) =e^{\frac{\left\vert
\alpha \right\vert ^{2}}{2}}e^{\alpha ^{\ast }a}e^{-\alpha a^{\dag }}$ and $%
\left\langle n\right\vert =\left\langle 0\right\vert \frac{1}{\sqrt{n!}}%
\frac{d^{n}}{ds^{n}}\exp \left( sa\right) |_{s=0}$ as well as $\left\vert
n\right\rangle \frac{1}{\sqrt{n!}}\frac{d^{n}}{dh^{n}}\exp \left( ha^{\dag
}\right) |_{h=0}\left\vert 0\right\rangle $. Here, we remain the
differential form. Thus, once the input CF is known, then the output PND can
be obtained by performing the integration in Eq. (\ref{8}).

\subsection{Wigner function}

The WF is a well-known quantum mechanical quasi-distribution function used
extensively in molecular and quantum optical computations as well as in
various theoretical contexts \cite{13}, whose negativity is a witness of the
nonclassicality of a quantum state \cite{14}. For a single-mode density
operator $\rho $, the WF in the coherent state representation $\left\vert
z\right\rangle $ can be expressed as
\begin{equation}
W(\varepsilon )=\frac{2e^{2\left\vert \varepsilon \right\vert ^{2}}}{\pi }%
\int \frac{d^{2}z}{\pi }\left\langle -z\right\vert \rho \left\vert
z\right\rangle \exp \left( 2\varepsilon z^{\ast }-2z\varepsilon ^{\ast
}\right) ^{\allowbreak },  \label{9}
\end{equation}%
with $\varepsilon =\left( x+ip\right) /\sqrt{2}$. Thus, we have the WF%
\begin{equation}
W_{in}(\varepsilon )=\frac{1}{\pi }\int \frac{d^{2}\alpha }{\pi }\chi
_{in}\left( \alpha \right) \exp \left( -\alpha \varepsilon ^{\ast
}+\varepsilon \alpha ^{\ast }\right) ,  \label{10}
\end{equation}%
for the input state $\rho _{in}$ and%
\begin{eqnarray}
W_{out}(\varepsilon ) &=&\frac{1}{\pi }\frac{1}{t^{2}}\int \frac{d^{2}\alpha
}{\pi }\chi _{in}\left( \alpha \right) \exp \left( -\frac{1}{t}\alpha
\allowbreak \varepsilon ^{\ast }+\frac{1}{t}\varepsilon \alpha ^{\ast
}\right)  \notag \\
&&\times \exp \left( -X\alpha \alpha ^{\ast }+Y\alpha ^{2}+Y\alpha ^{\ast
}{}^{2}\right) ,  \label{11}
\end{eqnarray}%
for the output state $\rho _{out}$. Next, we try our best to find the
input-output relationship of the WF.

Recalling the following identity \cite{15,16}%
\begin{align}
& \int \frac{d^{2}z}{\pi }\exp \left( \zeta \left\vert z\right\vert ^{2}+\xi
z+\eta z^{\ast }+fz^{2}+gz^{\ast 2}\right)  \notag \\
& =\frac{1}{\sqrt{\zeta ^{2}-4fg}}\exp \left( \frac{1}{\zeta ^{2}-4fg}\left(
-\zeta \xi \eta +\xi ^{2}g+\eta ^{2}f\right) \right) ,  \label{12}
\end{align}%
we find%
\begin{eqnarray}
&&\exp \left( -X\alpha \alpha ^{\ast }+Y\alpha ^{2}+Y\alpha ^{\ast
}{}^{2}\right)  \notag \\
&=&\sqrt{\frac{1}{X^{2}-4Y^{2}}}\int \frac{d^{2}\gamma }{\pi }\exp \left( -%
\frac{X}{X^{2}-4Y^{2}}\left\vert \gamma \right\vert ^{2}\right)  \notag \\
&&\times \exp \left( +i\alpha ^{\ast }\gamma +i\alpha \gamma ^{\ast }-\frac{Y%
}{X^{2}-4Y^{2}}\gamma ^{2}-\frac{Y}{X^{2}-4Y^{2}}\gamma ^{\ast 2}\right) .
\label{13}
\end{eqnarray}%
Then we change Eq.(\ref{13}) into%
\begin{eqnarray}
W_{out}(\varepsilon ) &=&\sqrt{\frac{1}{X^{2}-4Y^{2}}}\frac{1}{t^{2}}\int
\frac{d^{2}\gamma }{\pi }W_{in}(\frac{1}{t}\varepsilon +i\gamma )  \notag \\
&&\times \exp \left( -\frac{X}{X^{2}-4Y^{2}}\left\vert \gamma \right\vert
^{2}-\frac{Y}{X^{2}-4Y^{2}}\gamma ^{2}-\frac{Y}{X^{2}-4Y^{2}}\gamma ^{\ast
2}\right) ,  \label{14}
\end{eqnarray}%
that is,
\begin{eqnarray}
W_{out}(\varepsilon ) &=&\frac{\allowbreak 2}{T}\int \frac{d^{2}\gamma }{\pi
}W_{in}(\frac{1}{t}\varepsilon +i\gamma )  \notag \\
&&\times \exp \left( -\frac{2t^{2}\cosh 2\lambda }{T}\left\vert \gamma
\right\vert ^{2}-\frac{t^{2}\sinh 2\lambda }{T}\gamma ^{2}-\frac{t^{2}\sinh
2\lambda }{T}\gamma ^{\ast 2}\right) ,  \label{15}
\end{eqnarray}%
with $T=r^{2}\left( 2\bar{n}+1\right) $.

Replacing $z=\frac{1}{t}\varepsilon +i\gamma $, then we change Eq.(\ref{15})
in another form as follows%
\begin{eqnarray}
W_{out}(\varepsilon ) &=&\allowbreak \frac{2}{T}\int \frac{d^{2}z}{\pi }%
W_{in}(z)\exp \left( -\frac{2\cosh 2\lambda }{T}\left\vert zt-\varepsilon
\right\vert ^{2}\right)  \notag \\
&&\times \exp \left( \frac{\sinh 2\lambda }{T}\left( zt-\varepsilon \right)
^{2}+\frac{\sinh 2\lambda }{T}\left( z^{\ast }t-\varepsilon ^{\ast }\right)
^{2}\right) .  \label{16}
\end{eqnarray}%
Thus the transformation formula of the input-output WF in the BS formalism
have been obtained. Once the input WF is known, then the output WF can be
obtained by performing the integration in Eq. (\ref{15}) or (\ref{16}).

\section{Optical-channel formalism}

Now, we come back to the realistic situation. Noticing the correspondences
between the BS formalism and the optical-channel formalism, i.e., $%
t=e^{-\kappa \tau /2},r=\sqrt{\left( 1-e^{-\kappa \tau }\right) }$, we can
easily obtain the time evolution formulas of the density operator, the PND
and the WF at the moment $\tau $ from Eqs.(\ref{7}), (\ref{8}) and (\ref{16}%
).

\subsection{Density operator}

From Eq.(\ref{7}), we obtain the density operator of the optical field at
any time
\begin{eqnarray}
\rho \left( \tau \right) &=&\int \frac{d^{2}\alpha }{\pi }\chi \left( \alpha
e^{-\kappa \tau /2};0\right) D_{a}\left( -\alpha \right) \exp \left(
-M\left( 1-e^{-\kappa \tau }\right) \left\vert \alpha \right\vert ^{2}\right)
\notag \\
&&\times \exp \left( N\left( 1-e^{-\kappa \tau }\right) \alpha ^{2}+N\left(
1-e^{-\kappa \tau }\right) \alpha ^{\ast 2}\right) .  \label{17}
\end{eqnarray}%
where $\chi \left( \alpha ;0\right) $\ is the CF of the initial state $\rho
\left( 0\right) $. Hence, as long as we know the initial CF, we can obtain
the density operator at any time.

\subsection{Photon number distribution}

From Eq.(\ref{8}), we obtain the PND of the optical field at any time%
\begin{eqnarray}
P\left( n;\tau \right) &=&\frac{1}{n!}\frac{d^{2n}}{dh^{n}ds^{n}}\int \frac{%
d^{2}\alpha }{\pi }\chi \left( \alpha e^{-\kappa \tau /2};0\right) \exp
\left( -\left( M\left( 1-e^{-\kappa \tau }\right) +\frac{1}{2}\right)
\left\vert \alpha \right\vert ^{2}\right)  \notag \\
&&\times \exp \left( N\left( 1-e^{-\kappa \tau }\right) \alpha ^{2}+N\left(
1-e^{-\kappa \tau }\right) \alpha ^{\ast 2}+\allowbreak h\alpha ^{\ast
}-s\alpha +hs\right) |_{s=h=0}.  \label{18}
\end{eqnarray}%
One can verify the correctness of this formula in two extreme cases, i.e. $%
\tau \rightarrow 0$ and $\tau \rightarrow \infty $. Once the initial CF is
known, then the final PND can be obtained by performing the integration in
Eq. (\ref{18}).

\subsection{Wigner function}

Using the correspondences $W_{in}(z)\rightarrow W(z;0)$ and $%
W_{out}(z)\rightarrow W(\varepsilon ;\tau )$, Eq.(\ref{16}) can further
expressed as%
\begin{eqnarray}
W(\varepsilon ;\tau ) &=&\allowbreak \frac{2}{T}\int \frac{d^{2}z}{\pi }%
W(z;0)\exp \left( -\frac{2\cosh 2\lambda }{T}\left\vert ze^{-\kappa \tau
/2}-\varepsilon \right\vert ^{2}\right)  \notag \\
&&\times \exp \left( \frac{\sinh 2\lambda }{T}\left( ze^{-\kappa \tau
/2}-\varepsilon \right) ^{2}+\frac{\sinh 2\lambda }{T}\left( z^{\ast
}e^{-\kappa \tau /2}-\varepsilon ^{\ast }\right) ^{2}\right) ,  \label{19}
\end{eqnarray}%
with $T=\left( 2\bar{n}+1\right) \left( 1-e^{-\kappa \tau }\right) $. This
is an important result, which shows the evolution formula of the WF for the
optical field through the optical channel. Thus once the initial WF is
known, the WF at any time can be obtained by performing the integration in
Eq. (\ref{19}). In particular, when $\lambda =0$, $\bar{n}=0$, Eq.(\ref{19})
is reduced to the result in Ref.\cite{17}. Moreover, when $\kappa \tau
\rightarrow \infty $, $W(\varepsilon ;\tau )\rightarrow W_{\rho
_{b}}(\varepsilon )$, just like in Appendix A.

\section{Application of the evolution formulas}

Actually, we have derive some general formulas in the general Gaussian
reservoir (or environment) with squeezed thermal state. However, for the
sake of simplicity, we discuss the time evolution of the PND and the WF by
taking thermal channel (that is, we set $\lambda =0$ and $M=\bar{n}+\frac{1}{%
2},N=0$) and the initial single-photon-added coherent state%
\begin{equation}
\left\vert \psi _{a}\right\rangle =\frac{1}{\sqrt{\Gamma }}a^{\dag
}\left\vert \varkappa \right\rangle .  \label{20}
\end{equation}%
Here $\left\vert \varkappa \right\rangle $\ is a coherent state and $\Gamma
=1+\left\vert \varkappa \right\vert ^{2}$ is the normalization factor.

\subsection{Photon number distribution}

Using Eq.(\ref{B.3}) and (\ref{18}), we have%
\begin{eqnarray}
P\left( n;\tau \right) &=&\frac{\Lambda }{n!\Gamma }\frac{d^{2+2n}}{%
dh_{1}ds_{1}dh^{n}ds^{n}}\exp \left( +\allowbreak \varkappa
h_{1}+\allowbreak s_{1}\varkappa ^{\ast }+h_{1}s_{1}+hs\right)  \notag \\
&&\times \exp \left( \Lambda \left( \allowbreak h_{1}e^{-\kappa \tau
/2}+\varkappa ^{\ast }e^{-\kappa \tau /2}-s\right) \left( h-\varkappa
e^{-\kappa \tau /2}-s_{1}\allowbreak e^{-\kappa \tau /2}\right) \right)
|_{s=h=s_{1}=h_{1}=0},  \label{21}
\end{eqnarray}%
with $\Lambda =\left( \allowbreak \bar{n}-\bar{n}e^{-\kappa \tau }+1\right)
^{-1}$. After verification, we find that $P\left( n;0\right) \rightarrow
P_{\left\vert \psi _{a}\right\rangle }\left( n\right) $ and $P\left(
n;\infty \right) \rightarrow \bar{n}^{n}/\left( \bar{n}+1\right) ^{n+1}$ as
expected.

In Fig.2, we plot the PND for two different cases with (a) $\varkappa
=0.5+0.5i$ $\bar{n}=0.5$ and (b) $\varkappa =1$, $\bar{n}=0.3$ at different
evolution time. The blue, purple, brown and green bars are corresponding to
the cases at $\kappa \tau =0$, $\kappa \tau =0.05$, $\kappa \tau =0.25$, and
$\kappa \tau \rightarrow \infty $. Taking $P\left( 0\right) $ as an example,
as shown for the first bars in Fig.2 (a), the zero-photon component $P\left(
0\right) $ is increasing with the evolution time $\kappa \tau $, where we
find $P\left( 0\right) =$0 at $\kappa \tau =0$, $P\left( 0\right) =$%
0.0302572 at $\kappa \tau =0.05$, $P\left( 0\right) =$0.145164 at $\kappa
\tau =0.25$, and $P\left( 0\right) =$0.666667 at $\kappa \tau \rightarrow
\infty $.
\begin{figure}[tbp]
\label{Fig2} \centering\includegraphics[width=0.9\columnwidth]{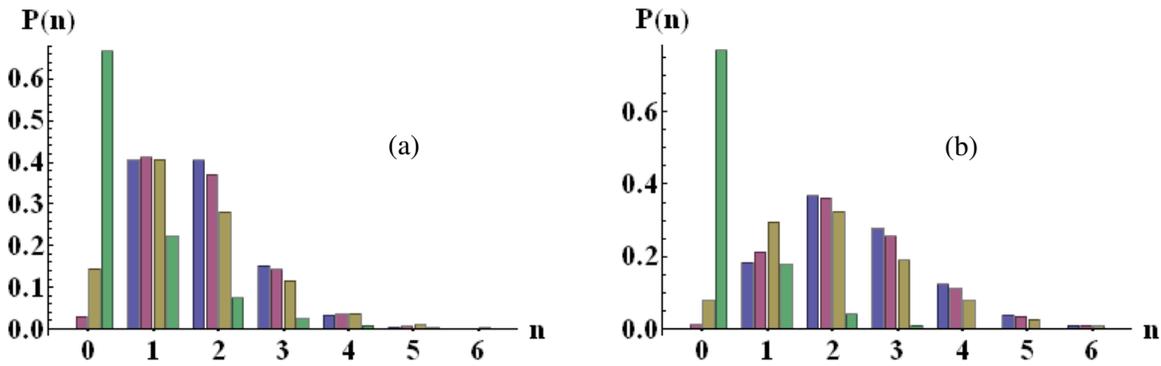}
\caption{(Colour online) Time evolution of the photon number distribution
for a single-photon-addition coherent state as the initial state through
thermal channel with (a) $\varkappa =0.5+0.5i$ $\bar{n}=0.5$; (b) $\varkappa
=1$, $\bar{n}=0.3$. The blue, purple, brass and green bars are corresponding
to the cases $\protect\kappa \protect\tau =0$, $\protect\kappa \protect\tau %
=0.05$, $\protect\kappa \protect\tau =0.25$, and $\protect\kappa \protect%
\tau \rightarrow \infty $ (i.e. the thermal state).}
\end{figure}

\subsection{Wigner function}

Substituting Eq.(\ref{B.4}) into Eq.(\ref{19}), we have%
\begin{equation}
W(\varepsilon ;\tau )=\frac{2}{\pi \Gamma }\left( g_{1}\left\vert \varkappa
\right\vert ^{2}-2g_{2}\varkappa \varepsilon ^{\ast }-2g_{2}\varkappa ^{\ast
}\varepsilon +4g_{3}\left\vert \varepsilon \right\vert ^{2}-g_{4}\right)
\exp \left( -2g\left\vert \varkappa e^{-\kappa \tau /2}-\varepsilon
\right\vert ^{2}\right) ,  \label{22}
\end{equation}%
with $g_{1}=\left( 2ge^{-\kappa \tau }-1\right) ^{2}g$, $g_{2}=\allowbreak
2g^{3}e^{-3\kappa \tau /2}-g^{2}e^{-\kappa \tau /2}$, $g_{3}=g^{3}e^{-\kappa
\tau }$, $g_{4}=2g^{2}e^{-\kappa \tau }-g$, and $g=\left( 2\bar{n}-2\bar{n}%
e^{-\kappa \tau }+1\right) ^{-1}$. In particular, when $\kappa \tau
\rightarrow 0$, we find that $W(\varepsilon ;0)\rightarrow W_{\left\vert
\psi _{a}\right\rangle }(\varepsilon )$, (see appendix B (3)); while $\kappa
\tau \rightarrow \infty $, we have $W(\varepsilon ;\infty )=\frac{2}{\pi }%
\frac{1}{2\bar{n}+1}\exp \left( -\frac{2}{2\bar{n}+1}\left\vert \varepsilon
\right\vert ^{2}\right) $, (see appendix A (3) with $\lambda =0$) which is
just the WF of the thermal state, also the environment, as expected.

In Fig.3, we plot the WFs for the case $\varkappa =0.5+0.5i$, $\bar{n}=0.5$
at $\kappa \tau =0$, $\kappa \tau =0.05$, $\kappa \tau =0.25$, and $\kappa
\tau \rightarrow \infty $. With the increase of time, the negativity of the
WF gradually decreases until it disappears.
\begin{figure}[tbp]
\label{Fig3} \centering\includegraphics[width=0.9\columnwidth]{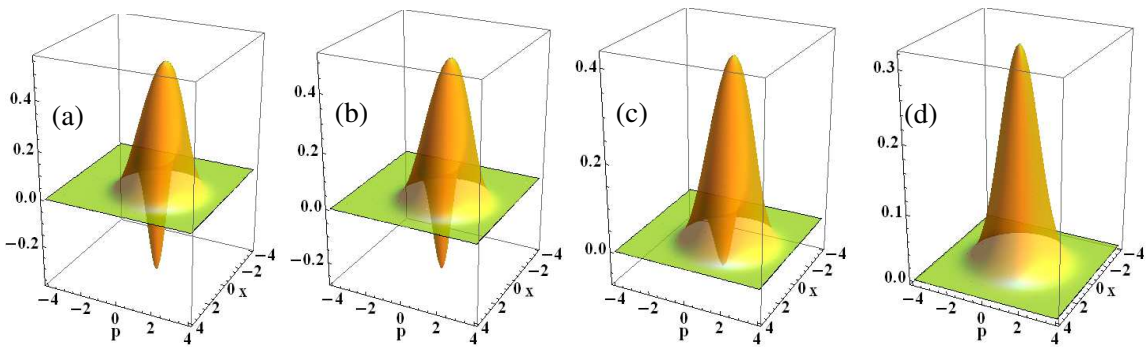}
\caption{(Colour online) Time evolution of the Wigner function for a
single-photon-addition coherent state with $\varkappa =0.5+0.5i$ as the
initial state in thermal channel with $\bar{n}=0.5$. (a) $\protect\kappa
\protect\tau =0$; (b) $\protect\kappa \protect\tau =0.05$; (c) $\protect%
\kappa \protect\tau =0.25$; (d) $\protect\kappa \protect\tau \rightarrow
\infty $ (i.e. the thermal state).}
\end{figure}

\section{Conclusion}

In this paper, we have explored the problem on the optical fields
propagation in realistic environments.\ For our purpose, we have abstracted
it into the optical-channel formalism and the BS formalism. In the BS
formalism, we derive the input-output relation of the density operator in
the CF formalism, which is convenient for obtaining the formulas of the PND
and the WF. Using the correspondences of these two formalisms, we easily
obtain the time evolution formulas of the density operator, the PND and the
WF. These formulas are very important in quantum optics and quantum
statistics. As long as we know the CF or the WF of the initial optical
field, we can readily find the statistical properties of the optical fields
at any time. As an example, we have discussed the case of the
single-photon-added coherent state propagated in thermal channel.

\begin{acknowledgments}
This work was supported by the Natural Science Foundation of Jiangxi
Province of China (20151BAB202013) and the Research Foundation of the
Education Department of Jiangxi Province of China (GJJ150338).
\end{acknowledgments}

\textbf{Appendix A: Some character of squeezed thermal state}

In this appendix, we give a good expression of the density operator and then
calculate the CF and WF for squeezed thermal state.

1) Density operator

Recalling the P-function of squeezed thermal state $\rho _{b}$%
\begin{equation}
\rho _{b}=\frac{1}{\bar{n}}\int \frac{d^{2}\eta }{\pi }e^{-\frac{1}{\bar{n}}%
\left\vert \eta \right\vert ^{2}}S\left( \lambda \right) \left\vert \eta
\right\rangle \left\langle \eta \right\vert S^{\dag }\left( \lambda \right) .
\label{A.1}
\end{equation}%
and noticing the squeezing operator $S\left( \lambda \right) =\int \frac{dx}{%
\sqrt{e^{-\lambda }}}\left\vert \frac{x}{e^{-\lambda }}\right\rangle
\left\langle x\right\vert $ with $\left\vert x\right\rangle =\pi ^{-\frac{1}{%
4}}e^{-x^{2}/2+\sqrt{2}xb^{\dag }-b^{\dag 2}/2}\left\vert 0\right\rangle $,
leading to%
\begin{equation}
S\left( \lambda \right) \left\vert \eta \right\rangle =\frac{1}{\sqrt{\mu }}%
\exp \left( -\frac{\left\vert \eta \right\vert ^{2}}{2}-\frac{\nu }{2}\eta
^{2}+\frac{\eta }{\mu }b^{\dag }+\frac{\nu }{2}b^{\dag 2}\right) \left\vert
0\right\rangle ,  \label{A.2}
\end{equation}%
with $\mu =\cosh \lambda $ and $\nu =\tanh \lambda $, we know%
\begin{eqnarray}
\rho _{b} &=&\frac{1}{\mu \bar{n}}\int \frac{d^{2}\eta }{\pi }\exp \left( -%
\frac{1}{\bar{n}}\left\vert \eta \right\vert ^{2}-\left\vert \eta
\right\vert ^{2}-\frac{\nu }{2}\eta ^{2}-\frac{\nu }{2}\eta ^{\ast 2}\right)
\notag \\
&&\times \exp \left( \frac{1}{\mu }\eta b^{\dag }+\frac{\nu }{2}b^{\dag
2}\right) \left\vert 0\right\rangle \left\langle 0\right\vert \exp \left(
\frac{1}{\mu }\eta ^{\ast }b+\frac{\nu }{2}b^{2}\right) .  \label{A.3}
\end{eqnarray}%
This expression of the density operator can help us calculate its
statistical properties.

2) Characteristic function

The CF of squeezed thermal state is given by $\chi _{b}\left( \beta \right)
= $Tr$\left( \rho _{b}D_{b}\left( \beta \right) \right) $ with $D_{b}\left(
\beta \right) =e^{\left\vert \beta \right\vert ^{2}/2}e^{-\beta ^{\ast
}b}e^{\beta b^{\dag }}$ \cite{18}. Thus, we have%
\begin{eqnarray}
\chi _{b}\left( \beta \right) &=&\frac{1}{\mu \bar{n}}e^{\frac{\left\vert
\beta \right\vert ^{2}}{2}}\int \frac{d^{2}\eta }{\pi }\exp \left( -\frac{1}{%
\bar{n}}\left\vert \eta \right\vert ^{2}-\left\vert \eta \right\vert ^{2}-%
\frac{\nu }{2}\eta ^{2}-\frac{\nu }{2}\eta ^{\ast 2}\right)  \notag \\
&&\left\langle 0\right\vert \exp \left( \frac{1}{\mu }\eta ^{\ast }b-\beta
^{\ast }b+\frac{\nu }{2}b^{2}\right) \exp \left( \frac{1}{\mu }\eta b^{\dag
}+\beta b^{\dag }+\frac{\nu }{2}b^{\dag 2}\right) \left\vert 0\right\rangle
\notag \\
&=&\exp \left( -M\left\vert \beta \right\vert ^{2}+N\beta ^{2}+N\beta ^{\ast
}{}^{2}\right) .  \label{A.4}
\end{eqnarray}

3) Wigner function

Using Eqs.(\ref{9}) and (\ref{A.3}), we obtain the WF of the squeezed
thermal state as follows%
\begin{equation}
W_{\rho _{b}}(\varepsilon )=\frac{2}{\pi }\frac{1}{2\bar{n}+1}\exp \left( -2%
\frac{\cosh 2\lambda }{2\bar{n}+1}\left\vert \varepsilon \right\vert ^{2}+%
\frac{\sinh 2\lambda }{2\bar{n}+1}\varepsilon ^{2}+\allowbreak \frac{\sinh
2\lambda }{2\bar{n}+1}(\varepsilon ^{\ast })^{2}\right) .  \label{A.5}
\end{equation}

\textbf{Appendix B: Some character of single-photon-added coherent state}

In this appendix, we derive the normalization factor, the CF and the WF for
the single-photon-added coherent state $\left\vert \psi _{a}\right\rangle =%
\frac{1}{\sqrt{\Gamma }}a^{\dag }\left\vert \varkappa \right\rangle $. The
density operator $\rho _{a}=\left\vert \psi _{a}\right\rangle \left\langle
\psi _{a}\right\vert $ can be expressed as%
\begin{equation}
\rho _{a}=\frac{1}{\Gamma }\exp \left( -\left\vert \varkappa \right\vert
^{2}\right) \frac{d^{2}}{dh_{1}ds_{1}}\exp \left( s_{1}a^{\dag }+\varkappa
a^{\dag }\right) \left\vert 0\right\rangle \left\langle 0\right\vert \exp
\left( h_{1}a+\varkappa ^{\ast }a\right) |_{s_{1}=h_{1}=0},  \label{B.1}
\end{equation}

1) Normalization factor

Using Tr$\left( \rho _{a}\right) =1$, we have the normalization factor%
\begin{eqnarray}
\Gamma &=&\exp \left( -\left\vert \varkappa \right\vert ^{2}\right)
\left\langle 0\right\vert \exp \left( h_{1}a+\varkappa ^{\ast }a\right) \exp
\left( s_{1}a^{\dag }+\varkappa a^{\dag }\right) \left\vert 0\right\rangle
|_{s_{1}=h_{1}=0}  \notag \\
&=&\exp \left( -\left\vert \varkappa \right\vert ^{2}\right) \frac{d^{2}}{%
dh_{1}ds_{1}}\exp \left( \left( h_{1}+\varkappa ^{\ast }\right) \left(
s_{1}+\varkappa \right) \right) |_{s_{1}=h_{1}=0}  \notag \\
&=&1+\left\vert \varkappa \right\vert ^{2},  \label{B.2}
\end{eqnarray}

2) characteristic function

Substituting $D_{a}\left( \alpha \right) =e^{\frac{\left\vert \alpha
\right\vert ^{2}}{2}}e^{-\alpha ^{\ast }a}e^{\alpha a^{\dag }}$ and Eq.(\ref%
{B.1}) into $\chi _{in}\left( \alpha \right) =$Tr$\left( \rho
_{a}D_{a}\left( \alpha \right) \right) $, we obtain the CF as follows%
\begin{eqnarray}
\chi _{in}\left( \alpha \right) &=&\frac{1}{\Gamma }\exp \left( \frac{%
\left\vert \alpha \right\vert ^{2}}{2}-\left\vert \varkappa \right\vert
^{2}\right) \frac{d^{2}}{dh_{1}ds_{1}}\exp \left( \left( h_{1}+\varkappa
^{\ast }-\alpha ^{\ast }\right) \left( s_{1}+\varkappa +\alpha \right)
\right) |_{s_{1}=h_{1}=0}  \notag \\
&=&\frac{1}{\Gamma }\left( \alpha \varkappa ^{\ast }-\left\vert \alpha
\right\vert ^{2}-\varkappa \alpha ^{\ast }+\left\vert \varkappa \right\vert
^{2}+1\right) \exp \left( -\frac{\left\vert \alpha \right\vert ^{2}}{2}%
+\alpha \varkappa ^{\ast }-\varkappa \alpha ^{\ast }\right) \allowbreak
\label{B.3}
\end{eqnarray}

3) Wigner function

Substituting Eq.(\ref{B.1}) into Eq.(\ref{9}), we obtain the WF
\begin{eqnarray}
W_{\left\vert \psi _{a}\right\rangle }(\varepsilon ) &=&\frac{2}{\pi \Gamma }%
\exp \left( 2\left\vert \varepsilon \right\vert ^{2}-\left\vert \varkappa
\right\vert ^{2}\right) \frac{d^{2}}{dh_{1}ds_{1}}\exp \left( \left(
h_{1}+\varkappa ^{\ast }-2\varepsilon ^{\ast }\right) \left( 2\varepsilon
-s_{1}-\varkappa \right) \right) ^{\allowbreak }|_{s_{1}=h_{1}=0}  \notag \\
&=&\frac{2}{\pi \Gamma }\left( \varkappa \varkappa ^{\ast }-2\varkappa
\varepsilon ^{\ast }-2\varepsilon \varkappa ^{\ast }+\allowbreak
4\varepsilon \varepsilon ^{\ast }-1\right) \exp \left( -2\left\vert
\varkappa -\varepsilon \right\vert ^{2}\right) .  \label{B.4}
\end{eqnarray}

\end{document}